# AMBIENT VIBRATIONS OF AGE-OLD MASONRY TOWERS: RESULTS OF LONG-TERM DYNAMIC MONITORING IN THE HISTORIC CENTRE OF LUCCA.

R.M. Azzara[1], M. Girardi[2], V. Iafolla[3], D. Lucchesi[3,2], C. Padovani[2], D. Pellegrini[2]

[1] Istituto Nazionale di Geofisica e Vulcanologia (INGV), Osservatorio Sismologico di Arezzo, Italy
[2] Institute of Information Science and Technologies "A. Faedo" CNR-ISTI, Pisa, Italy
[3] National Institute for Astrophysics, INAF-IAPS, Rome, Italy

## Abstract

The paper presents the results of an ambient vibration monitoring campaign conducted on so-called "Clock Tower" (Torre delle Ore), one the best known and most visited monuments in the historic centre of Lucca.

The vibrations of the tower were continuously monitored from November 2017 to March 2018 using high-sensitivity instrumentation. In particular, four seismic stations provided by the Istituto Nazionale di Geofisica e Vulcanologia and two three-axial accelerometers developed by AGI S.r.l., spin-off of the Istituto Nazionale di Astrofisica, were installed on the tower. The measured vibration level was generally very low, since the structure lies in the middle of a limited traffic area. Nevertheless, the availability of two different types of highly sensitive and accurate instruments allowed the authors to follow the dynamic behaviour of the tower during the entire monitoring period and has moreover provided cross-validation of the results.

## 1. Introduction

The first attempts to use ambient vibrations to characterize the dynamic properties of buildings date back to the 1970's (Trifunac, 1970). Nowadays, this technique has become standard practice (Celebi, 2017, Gallipoli et al., 2010, Kaya and Safak, 2015, Peeters et al. 1999, Prieto et al., 2010) thanks to the availability of sensitive instrumentation for measurement of low-amplitude vibrations, together with powerful algorithms and hardware for large dataset analysis. Ambient vibration monitoring is in fact very convenient, provided that sufficiently long records are available (Brincker, 2015), since it allows avoiding artificial vibration sources such as vibrodynes, thereby facilitating experiment management. Moreover, long-term ambient vibration measurements furnish important information on the sources of vibration, the influence of the environmental parameters on the structural dynamic properties and finally on the structural health status. Indeed, changes over time in a building's dynamic properties, such as modal properties (Doebling et al., 1996) or wave velocities (Todorovska, 2009), represent effective damage indicators, provided that the natural changes due to seasonal and daily environmental effects are taken into account. Long-term monitoring plays an important role when a structure is subjected to high vibration levels, such as those induced by traffic or construction sites (Wyjadlowski, 2017), or when the structure is located in high-seismicity areas (Celebi, 2017, Todorovska, 2009, Trifunac, 1970). More recently, the study of the dynamic effects on buildings of the surrounding environment has became a challenging research topic, involving disparate expertise from the fields of physics, geology and engineering, and sometimes referred to as "urban seismology" (Diaz et al., 2017), (Green et al., 2016), (Ritter et al. 2005). Moreover, the recent availability of low-cost measurement devices has encouraged the testing of large accelerometer networks for monitoring purposes (Barsocchi et al, 2018), (D'Alessandro et al., 2018), (Clementi et al., 2018), including through participative approaches as in (Matarazzo et al., 2018), where citizens were directly involved in monitoring the vibrations of bridges with their own smartphones (which are normally equipped with MEMS devices to measure accelerations).

With regard to the world's architectural heritage and monuments, the first examples of dynamic monitoring through ambient vibrations for damage detection purposes date back to (Gentile and Saisi, 2007), and (Ramos et al., 2010). A great deal of effort has been made to study the effects of traffic on ancient monuments, noteworthy amongst which is the work of (Pau and Vestroni, 2008) and (Bongiovanni et al., 2017) on dynamic characterization of the Colosseum, (Roselli et al, 2017) on other monuments in Rome, (Chiostrini et al., 1995), (Lacanna et al., 2016) and (Lacanna et al., 2019) in the historic centre of Florence, and (Erkal, 2017) on the traffic-induced vibrations on a minaret in Turkey. The availability of large datasets yielded by continuous monitoring of historic structures is however relatively recent. Some examples can be found in (Baraccani et al., 2017) on the Asinelli tower in Bologna, (Masciotta et al., 2017) on the Saint Torcato church, (Ubertini et al., 2017) on the ambient vibrations of the San Pietro bell tower in Perugia, (Lorenzoni et al., 2018) on post-earthquake vibration checks of some monumental buildings in l'Aquila, (Cabboi et al., 2017) on the continuous vibration monitoring of an age-old tower in Northern Italy, (Azzara et al., 2018) on the San Frediano bell tower in the historic centre of Lucca, and (Kita et al., 2019) on the Consoli Palace in Gubbio.

The present paper is aimed at investigating the dynamic behaviour of a medieval tower located in a historic centre and subjected to vibrations from the surrounding environment, the main goals being: to characterize the main sources of vibration, the trend of the tower's dynamic properties over time, and the tower's response to the activities in the historic centre. To this end, we rely on a dataset obtained from five months of continuous measurements (November 2017 to March 2018) on the so-called "Torre delle Ore" (henceforth Clock Tower) in the historic centre of Lucca. The measurements were recorded via two different sets of high-sensitivity instruments, both installed on the tower during the monitoring period: four three-axial seismometric stations (each equipped with a SL06 24-bit digitizer coupled to a SS20 electrodynamic velocity transducer) developed by SARA Electronic Instruments S.r.l. and furnished by the Istituto Nazionale di Geofisica e Vulcanologia (INGV), and two accelerometers (Iafolla et al., 2015), specifically adjusted for the experiment by Assist in Gravitation and Instrumentation (AGI S.r.l.) of Rome, a spin-off enterprise of the Istituto Nazionale di Astrofisica (INAF). The combined use of the two different kinds of measurement devices has allowed for a comparison of the instruments' performance, as well as cross-validation of the results obtained for the Clock Tower, which have also been corroborated by the outcomes of two-year continuous monitoring conducted on the nearby San Frediano bell tower (Azzara et a. 2018a, 2018b).

In addition, exploitation of the different characteristics of the seismometers and accelerometers has made it possible both to explore the dynamic behaviour of the tower and to highlight the effects of environmental vibrations in a wide frequency range. To the best of the authors' knowledge, the issues addressed in this paper are far from being fully investigated and the reported results constitute a novel contribution to understanding the dynamic response of the architectural heritage to anthropic and natural vibration sources.

## 2. The Clock Tower

Lucca was renowned in the past for the large number of towers in its skyline: of more than one hundred during Middle Ages, only ten or so of these fascinating monuments have survived until today (Figure 1). The Clock Tower is one of the best known and most visited age-old towers in Lucca, thanks to the peculiar shape of its bell chamber, which is clearly visible and recognizable throughout the entire historic centre. Built by local families (Concioni, 1988), since the last decade of the 15$^{th}$ century the Clock Tower has been used as a civic building, taking its name from the big clock visible on its southern facade (Figure 2). The Clock Tower rises at the corner between the roads named Via Arancio and Via Fillungo, one of the most popular in Lucca's historic centre; the adjacent buildings abut the tower on two sides for a height of about 13 m and constitute asymmetric boundary conditions for the tower's structure. The Clock Tower is 48.4 m high at the top of the bell chamber; it has a rectangular cross section of about 5.1 x 7.1 m and walls of thickness varying from about 1.77 m at the base to 0.85 m at the top. Two barrel vaults are set inside the tower at heights of about 12.5 and 42.3 m. The bell chamber, made up of four masonry pillars connected by elliptical arches, stands on the upper barrel vault and is covered by a pavilion roof constructed of wooden trusses and rafters. With regard to the materials constituting the masonry tower, visual inspection reveals that the masonry from the base up to a height of 15 m is made up of regular stone blocks and thin mortar joints. The upper walls are instead composed of regular stone blocks and bricks, also with thin joints. The pillars of the bell chamber are made of brick masonry. On 25 November 2016, during a preliminary experimental campaign, the ambient vibrations of the Clock Tower were monitored for a few hours via four SARA SS20 three-axial seismometric stations. The instruments were moved along the tower's height by adopting three different layouts and combining data in order to identify four natural vibration frequencies and mode shapes of the tower, as described in (Pellegrini et al, 2017). Data collected in this experiment also allowed evaluating the mechanical properties of the tower's constituent materials via model updating procedures, as described in (Girardi et al., 2019).

In November 2017 a long-term ambient vibration monitoring experiment was begun with the fitting of six instruments along the tower's height: four Sara SS20 seismometric stations (tri-axial velocimeters, Fig. 3) from the INGV, named in the following S.942, S.943, S.944, S.945, and two tri-axial accelerometers (Fig. 4) provided by the firm Assist in Gravitation and Instrumentation (AGI S.r.l.), named in the following S.1 and S.2. The INGV sensors were coupled with a 24-bit digitizer and the signal was sampled at 100 sps. The AGI data were acquired by a laptop at a sampling frequency of 20 Hz using a 24-bit digitizer. With regard to the main characteristics of the instruments used in the monitoring campaign, the SS20 velocity transducers have a nominal sensitivity of 200 V/m/s, eigenfrequency of 2 Hz, and usable band from 0.1 to 250 Hz. The AGI accelerometers have an acceleration noise density of $2 \cdot 10^{-8}$ g/$\sqrt{Hz}$ and usable band from $10^{-5}$ to 50 Hz. It is worth noting that together the instruments can cover a very wide band of frequencies.
The instrumentation was kept running on the tower up to March 2018. Figure 5 shows the layout of the sensors along the tower during the experiment: seismic station S.942 was placed at the base, collecting data from the ground, S.943 was at the height of 24 m above street level, the remaining two stations S.943 and S.944 were placed on the bell chamber, at about +42 m above street level. The AGI stations were both installed in the upper part of the structure, S.2 at +37 m and S.1 on the bell chamber level.

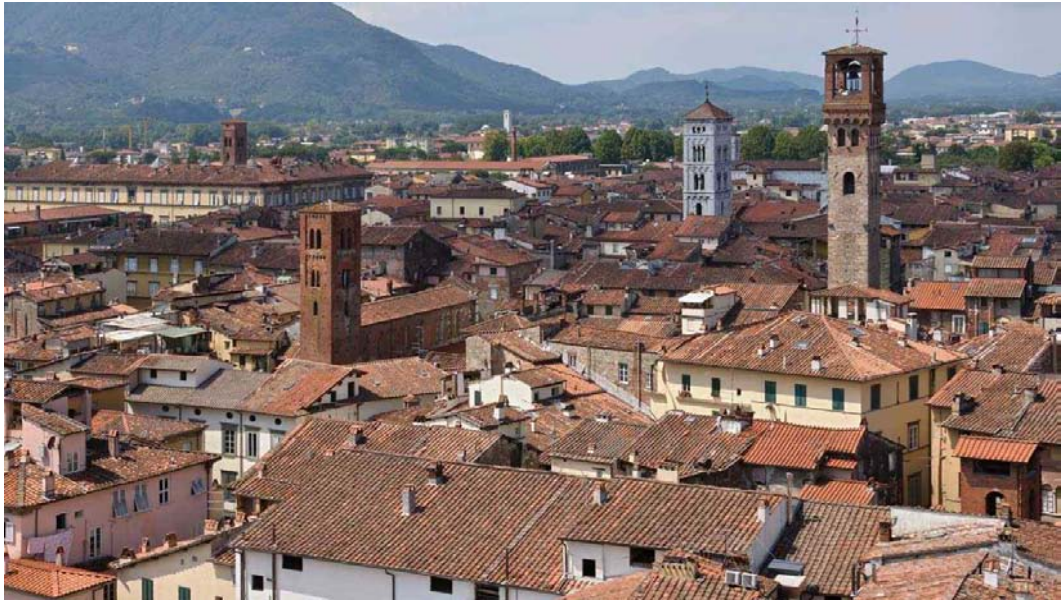

Figure 1. The historic centre of Lucca with the Clock Tower on the right.

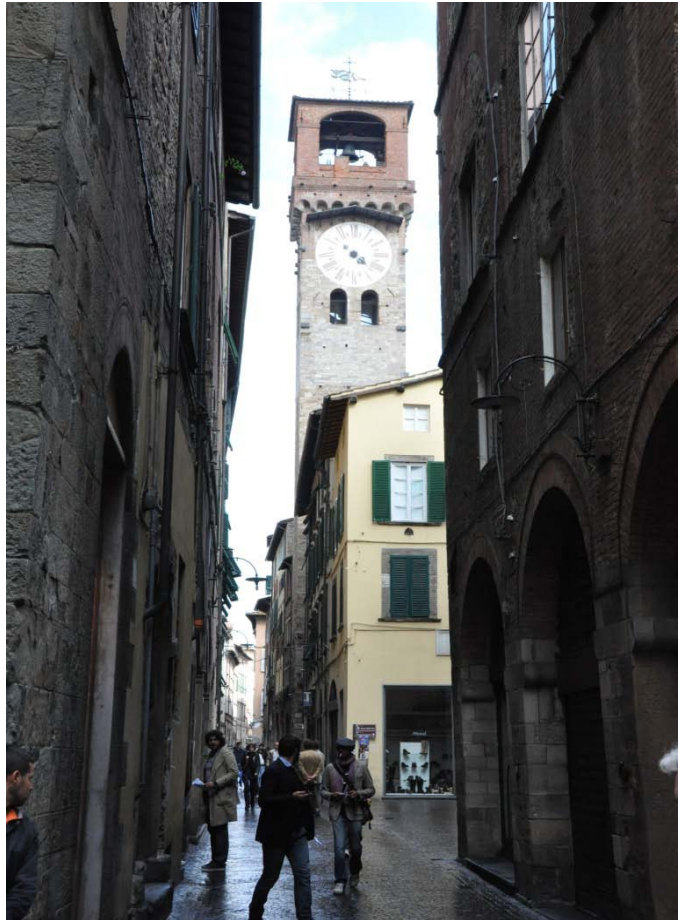
Figure 2. The Clock Tower: view from Via Fillungo.

The experiment was carried out in winter to take advantage of the tower's being closed to the public. Nevertheless, the experimental campaign encountered several problems, due to adverse weather conditions, a number of electrical blackouts, and the presence of many people working inside the tower and disrupting the data acquisition. In particular, the INGV seismic stations and the AGI sensor S.1 were removed on 8 December 2017 after an electrical shutdown. The velocimeters were installed again on 11 January 2018, while the AGI accelerometers were fitted to the tower on 11 February 2018. Moreover, from 29 January to 11 February 2018 all instrumentation was removed to allow a public ceremony to be held inside the tower. On 17 November 2017 a fifth INGV seismic station was installed inside a private flat in a building adjacent to the tower, and left active for some hours, with the aim of recording the natural frequencies of the building and evaluating its influence on the tower's dynamic behaviour. The data collected during the experiments are stored in a database hosted by the Mechanics of Materials and Structures Lab of CNR-ISTI in Pisa.

## 3. Data Analysis and results

All data have been analysed via the Covariance-driven Stochastic Subspace Identification method (SSI/Cov), an Operational Modal Analysis (OMA) technique in the time-domain implemented in the MACEC code (Reynders et al., 2016). To this end, the data have been divided into one-hour long datasets. Moreover, the data recorded by the INGV stations have also been processed via Experimental Modal Analysis (EMA), by applying the Complex Mode Indicator Function (CMIF) method (Shih et al., 1988) implemented in TruDI (Pellegrini, 2019) and considering the seismic station at the base (S942) as input signal. These last analyses were needed to take into account the composition of the signals at the base of the tower, whose spectrum shows a clear energy accumulation at about 3 Hz in the *x*-direction, along Via Fillungo. This same frequency was also recorded in the same direction by the station installed in the flat adjacent to the tower. In the authors' opinion, it represents a natural frequency of the building next to the tower and does not belong to the tower structure itself. Such conclusion is confirmed by the CMIF analysis, shown in Figure 6, where the three eigenvalues of the matrix $H[j\omega]^H H[j\omega]$ are reported vs. frequency, with $H[j\omega]$ the Frequency Response Function matrix of the system, estimated by the $H_1$ estimator, (Maia and Silva, 1997) and $\omega$ the circular frequency. Figure 6 refers to the analysis of a one-hour long dataset recorded by the INGV seismic stations: the four peaks represent the identified four frequencies of the tower, the 3 Hz frequency being much less accentuated.

The tower's frequencies are reported in Table 1, where the mean values (evaluated over the entire monitoring period from November 2017 to March 2018) of the four frequencies are reported, together with their relative differences (evaluated on the 1$^{st}$ and 99$^{th}$ percentile) for the two different instruments. The frequencies values measured on the Clock Tower are in good agreement with those of other towers with similar geometries (Bartoli et al., 2017). The dataset from the SS20 seismic stations includes 970 hours of recordings, while the AGI instrumentation yielded 1110 hours of tower measurements. The frequency values vs. time are reported in Figures 7 and 8 for the SS20 and the AGI instruments, respectively. The mean values evaluated by the two instruments are in very good agreement. The first and the second frequencies were detected by the two instruments in all records, and the variation in the two frequencies is on the order of 3%. This variation is also in good agreement with the findings of another long-term dynamic monitoring conducted on the San Frediano bell tower in Lucca (Azzara et al., 2018a). The third frequency was also detected by the seismic stations in almost all records, while the fourth frequency only appears when the excitation level in the tower's structure was increased (it was detected in about the 30% of the records), and its values appear to be widely dispersed. A fifth frequency at about 5.7 Hz was also detected in about 30% of the records. With regard to the modal damping ratios, reported in Table 2, they are widely dispersed with variations on the order of hundreds of percentage points. Their values vary from 4% for the highest frequency, to 0.5% for the first. The values yielded by the seismic stations are generally higher than those from the accelerometers.

The mode shapes of the tower corresponding to the four frequencies $f_1$, $f_2$, $f_3$, $f_4$ are shown in Figure 9; these have been extracted from the data recorded in November 2016 by taking advantage of the different layouts used for mapping the tower's vibrations. The first two frequencies refer to bending modes along the *x* (first) and *y* (second) directions, while the third is a torsional mode and the fourth is a torsional-bending mode involving movements of the structure's upper parts.

Figures 10 and 11 show further comparisons of the results obtained via the two instruments. In particular, Figure 10 shows the trend of the tower's fundamental frequency from 7 to 8 December 2017, extracted from the AGI (blue) and the SS20 (red) devices. The very good agreement between the two measurements indicates that the fluctuations exhibited in the frequency values do not depend on the instruments' functioning, but are rather due to environmental factors. Figure 11 shows a comparison of the hourly maximum accelerations (absolute values) recorded by the instruments during that same period. The differences between the amplitudes are mainly due to the different positions of the instruments, which were placed at slightly different levels along the tower. It is worth noting that maxima in the acceleration values correspond to minima in the frequency values. In fact, a look at the weather those days revealed very bad conditions, with strong wind velocities (see also Figure 15), thereby suggesting a connection between the frequency reduction shown in Figure 10 and the action of the wind, which due to the masonry's inability to withstand large tensile stresses, tends to decrease the tower's stiffness.

With regard to the influence of temperature on the tower's dynamic behaviour, Figure 12 shows the correlation of the first two natural frequencies with air temperature, measured by a sensor located in the Lucca Botanical Gardens, in Lucca's historic centre. The figure confirms the findings of other long-term vibration monitoring campaigns on historical towers (Cabboi et al., 2017), (Ubertini et al., 2017), including that carried out on the San Frediano bell tower in Lucca (Azzara et al., 2018a): frequencies tend to increase with temperature. This behaviour is confirmed by the results described in (Girardi et al., 2017), which reports on a nonlinear FE analysis of the Clock Tower conducted by modelling the effects of the thermal variations on the structure's frequencies. The numerical simulation supports the hypothesis that an increase in temperature induces a reduction in the fracture strains inside masonry, thus increasing the global stiffness of the structure.

With regard to negative temperature values, Figure 12 reveals opposite trend for frequencies, which turn out to increase when the temperature decreases. This phenomenon was mainly observed in the latter part of February 2018, when temperatures persistently decreased below zero, as highlighted by Figure 13. Correspondingly, the figure shows the change in the frequency trend. The behaviour below freezing has also been reported for similar weather conditions in (Ubertini et al., 2017) and (Cabboi et al., 2017). It is also worth noting that, as long as the temperature remains above zero, frequencies respond to temperature variations with a phase shift in time, while, below zero, the frequency and temperature appear to be in phase.

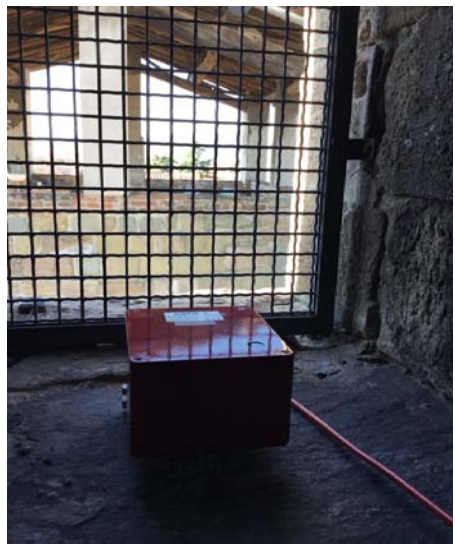

Figure 3. Sensor S.943 at level +24 m

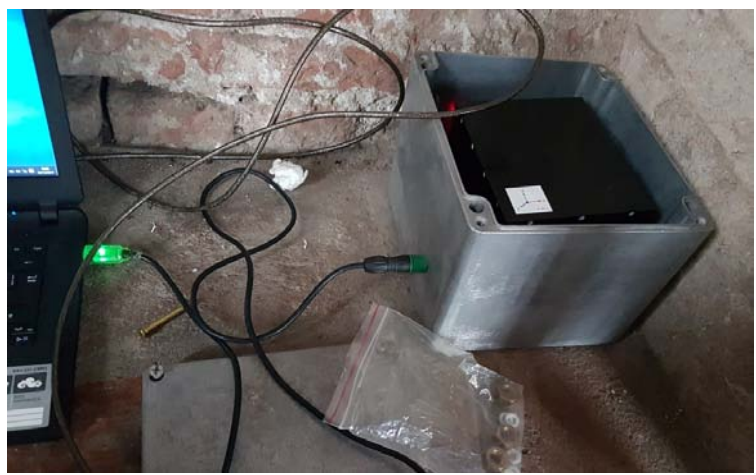

Figure 4. Sensor S.2 at level +37 m inside the tower.

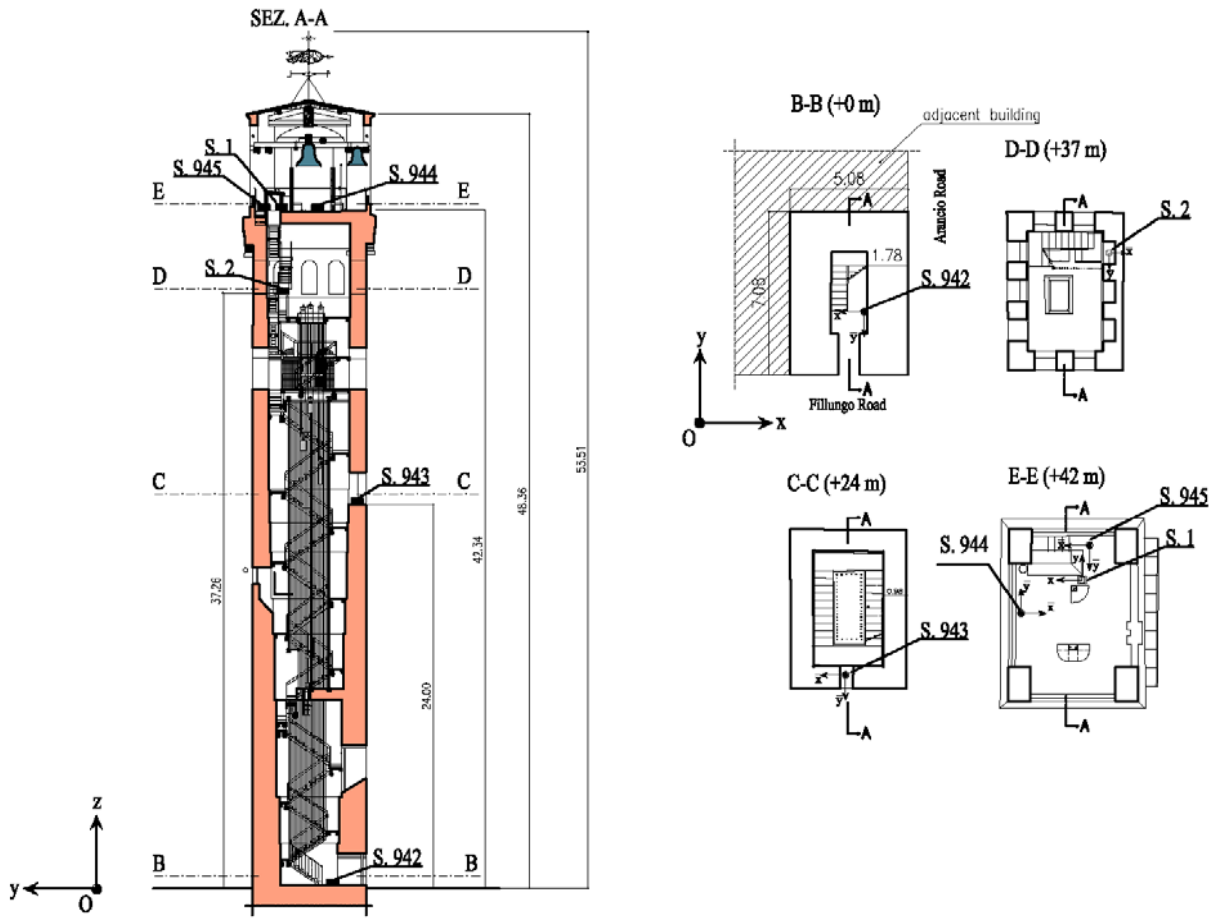

Figure 5. Sensor arrangement on the Clock tower: sensors S.942, S.943, S.944, S.945 (INGV), sensors S.1, S.2 (AGI).

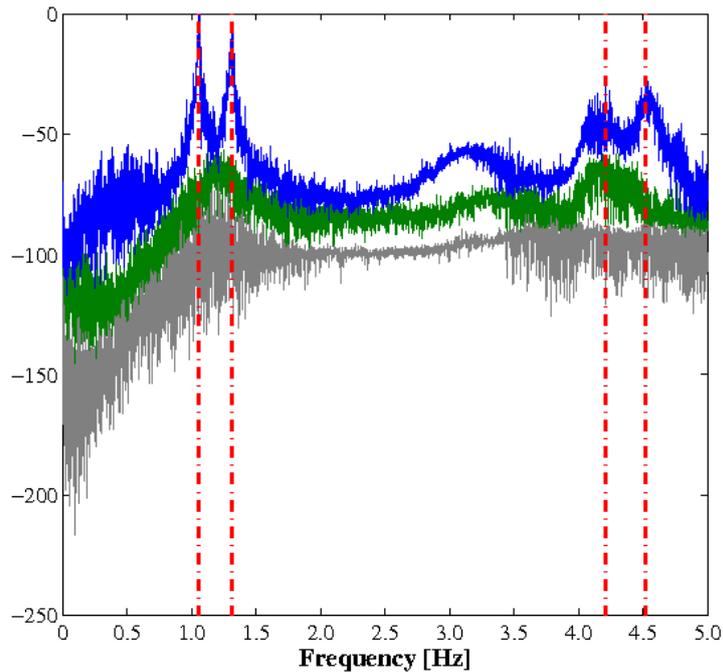

Figure 6. The three eigenvalues of the matrix $H[j\omega]^H H[j\omega]$ vs. the frequency on a log-magnitude scale.

|      | $f_1$ [Hz] | $\Delta_1$ [%] | $f_2$ [Hz] | $\Delta_2$ [%] | $f_3$ [Hz] | $\Delta_3$ [%] | $f_4$ [Hz] | $\Delta_4$ [%] |
|------|------------|----------------|------------|----------------|------------|----------------|------------|----------------|
| **SS20** | **1.0281** | 3.65 | **1.2813** | 3.40 | **4.0524** | 8.67 | **4.4858** | 10.68 |
| **AGI**  | **1.0318** | 2.94 | **1.2829** | 2.63 | **4.0712** | 7.26 | **4.4654** | 15.83 |

Table 1. Mean values $f_i$ of the natural frequencies measured by the two instruments during the monitoring period and their variation $\Delta_i = (f_i^1 - f_i^{99})/f_i^1$, where $f_i^1$ and $f_i^{99}$ represent the first and 99th percentile of the dataset, respectively.

|      | $\xi_1$ [%] | $\Delta_1$ [%] | $\xi_2$ [%] | $\Delta_2$ [%] | $\xi_3$ [%] | $\Delta_3$ [%] | $\xi_4$ [%] | $\Delta_4$ [%] |
|------|-------------|----------------|-------------|----------------|-------------|----------------|-------------|----------------|
| **SS20** | **0.90** | 165 | **1.26** | 150 | **2.03** | 450 | **1.88** | 885 |
| **AGI**  | **0.69** | 95  | **0.85** | 64  | **1.41** | 628 | **1.01** | 983 |

Table 2. Mean values $\xi_i$ of the modal damping ratios measured by the two instruments during the monitoring period and their variation $\Delta_i = (\xi_i^5 - \xi_i^{95})/\xi_i^5$, where $\xi_i^5$ and $\xi_i^{95}$ represent the fifth and 95th percentile of the dataset, respectively.

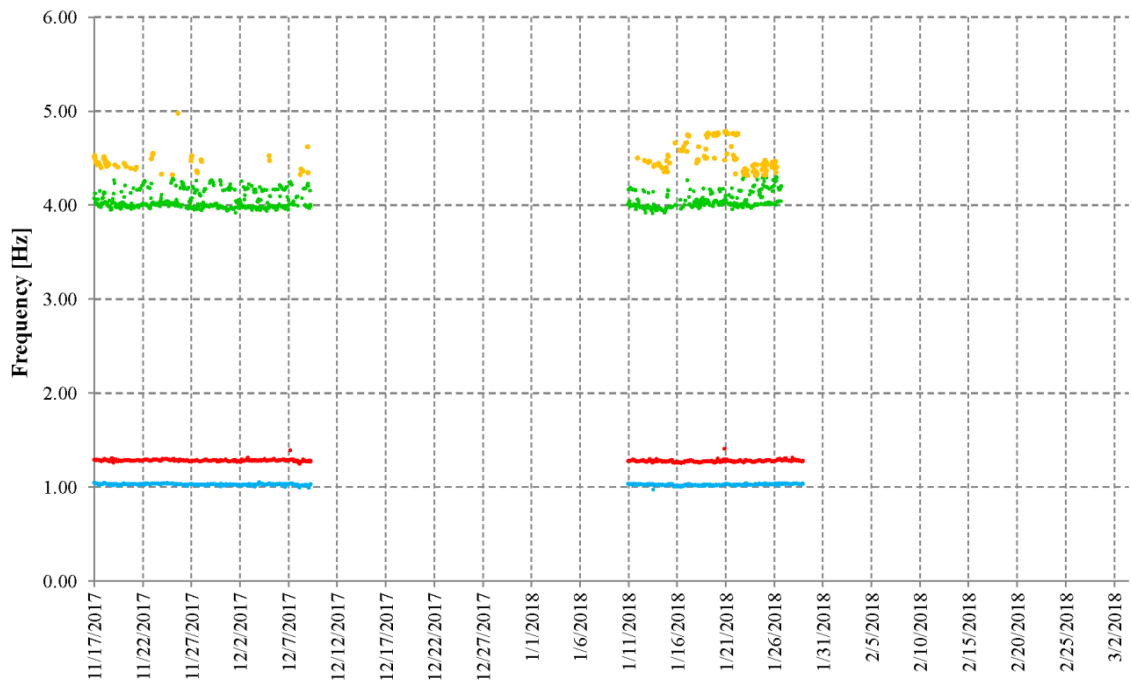

Figure 7. The tower's first four natural frequencies [Hz] detected by SS20 velocimeters during the monitoring period.

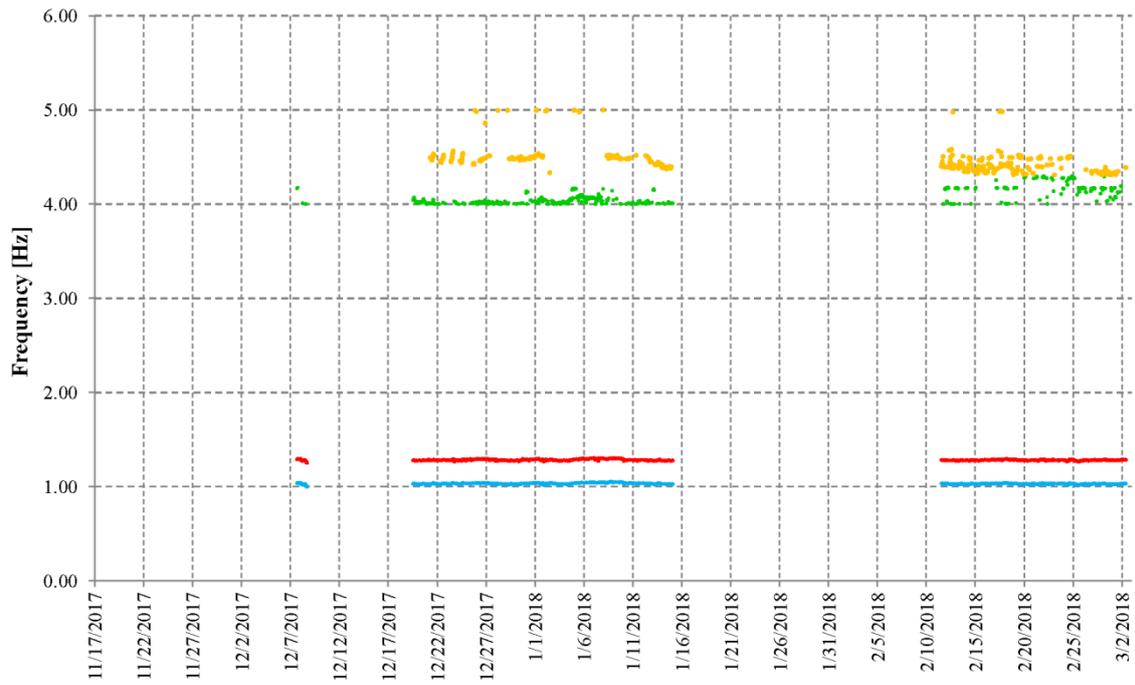

Figure 8. The tower's first four natural frequencies [Hz] detected by AGI accelerometers during the monitoring period.

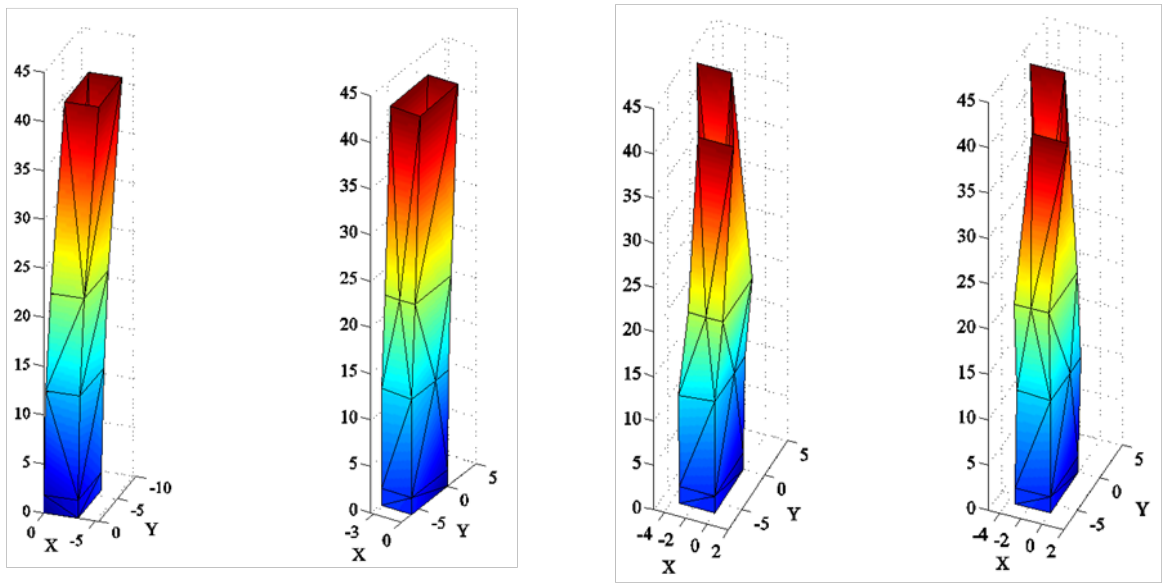

Figure 9. The tower's first four mode shapes [Hz] corresponding (from the left) to the frequencies $f_1$, $f_2$, $f_3$, $f_4$.

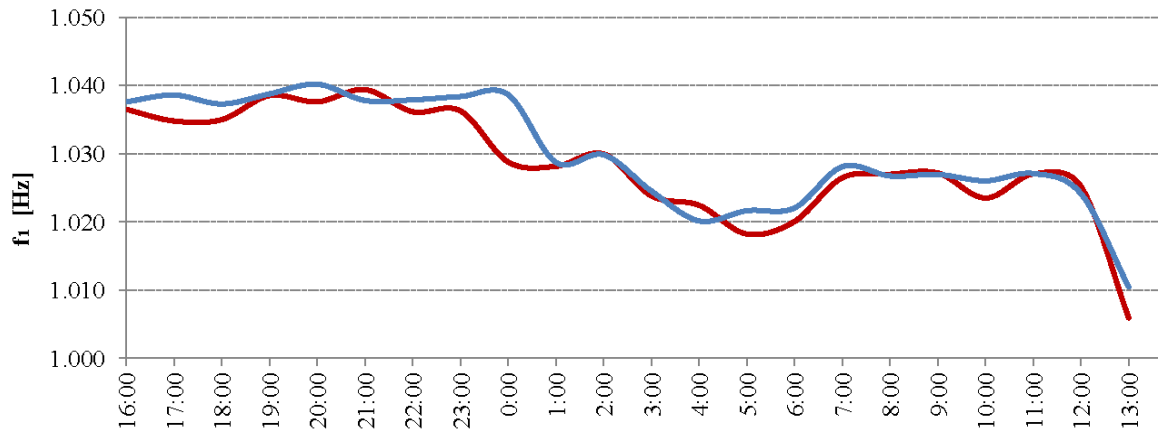

Figure 10. The tower's fundamental frequency obtained via AGI stations (blue) and SS20 stations (red) data from 7 to 8 December 2017 (respectively at 16:00 and 13:00).

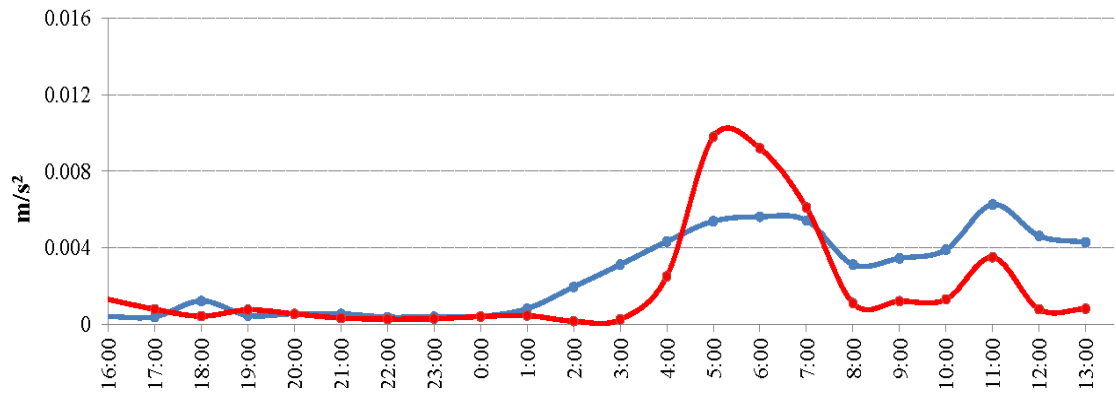

Figure 11. Maximum absolute accelerations recorded by the S.2 station (blue, +37 m) and the S. 945 station (red, +42 m) from 7 to 8 December 2017 (respectively at 16:00 and 13:00).

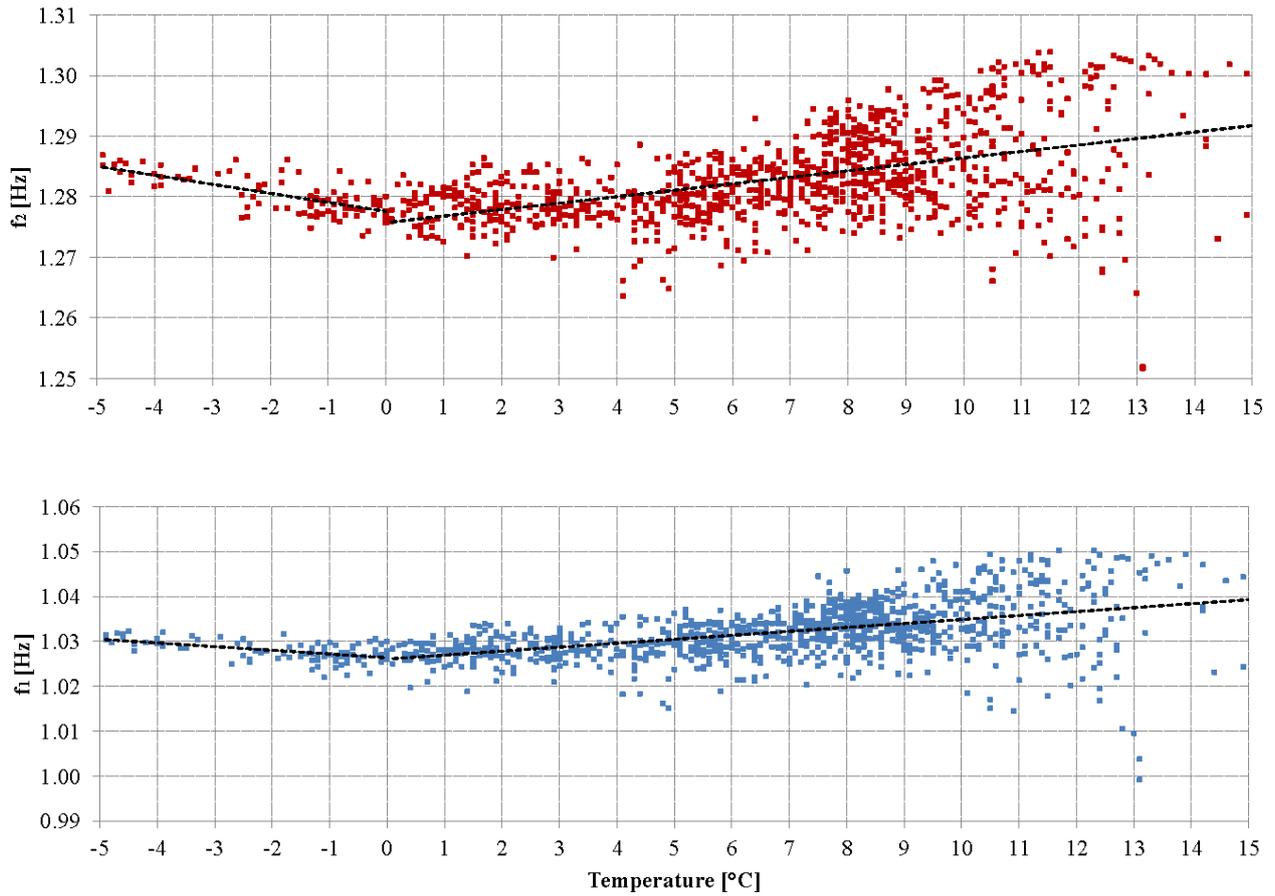

Figure 12. The first (blue) and second (red) natural frequencies of the tower [Hz] vs. temperature [°C].

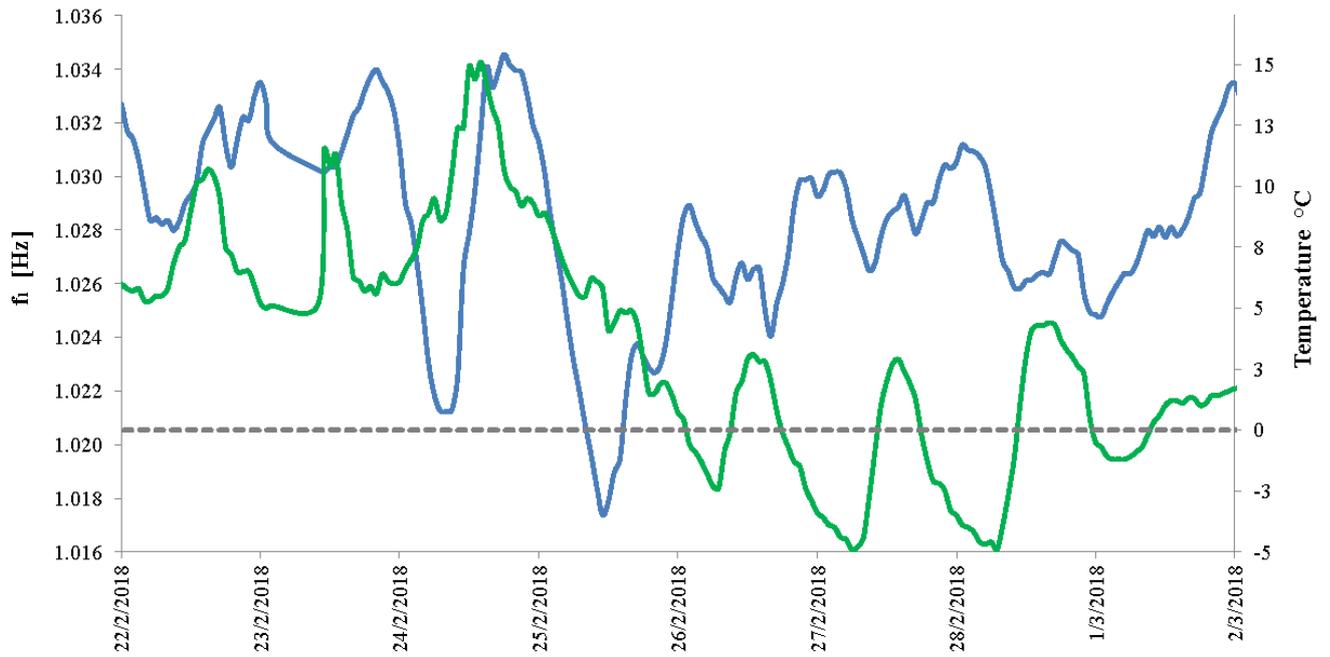

Figure 13. The tower's first (blue line) natural frequency [Hz] vs. temperature [°C] (green line), during February 2018.

The trend over time of the maximum accelerations recorded by AGI sensors is shown in Figure 14. The vibration level in the tower is very low, generally under $2 \cdot 10^{-3}$ m/s$^2$, thanks to restrictions to vehicular traffic in force in the Lucca historic centre. The figure highlights an increase in the tower's vibration level during the Christmas holiday period (from 22 December to 1 January), connected to the increased pedestrian traffic in the historic centre, with a peak on New Year's Day. The figure also reveals a systematic increase in the accelerations, on average of one order of magnitude, from 9 February onward. This is due to the ringing of the tower bells, which had been suspended for restoration of the ancient clock and was restarted on 16 January 2018. The bell system is composed of three bronze bells, fixed at their supports and rung by hammers each quarter an hour in the $x$ direction (along Via Fillungo). Indeed, Figure 14 shows significant amplification along $x$ (blue dots).

With regard to the environmental parameters that could influence the acceleration levels on the tower, Figure 15 plots the daily average values of the maximum hourly accelerations in the $x$ (blue) and $y$ (red) directions, for about a month, together with the daily maximum wind speed recorded at Pieve di Compito, about 10 km from the Lucca historic centre (data available at www.sir.toscana.it). The figure clearly shows a correspondence between the peaks in the acceleration levels and those in the wind speed.

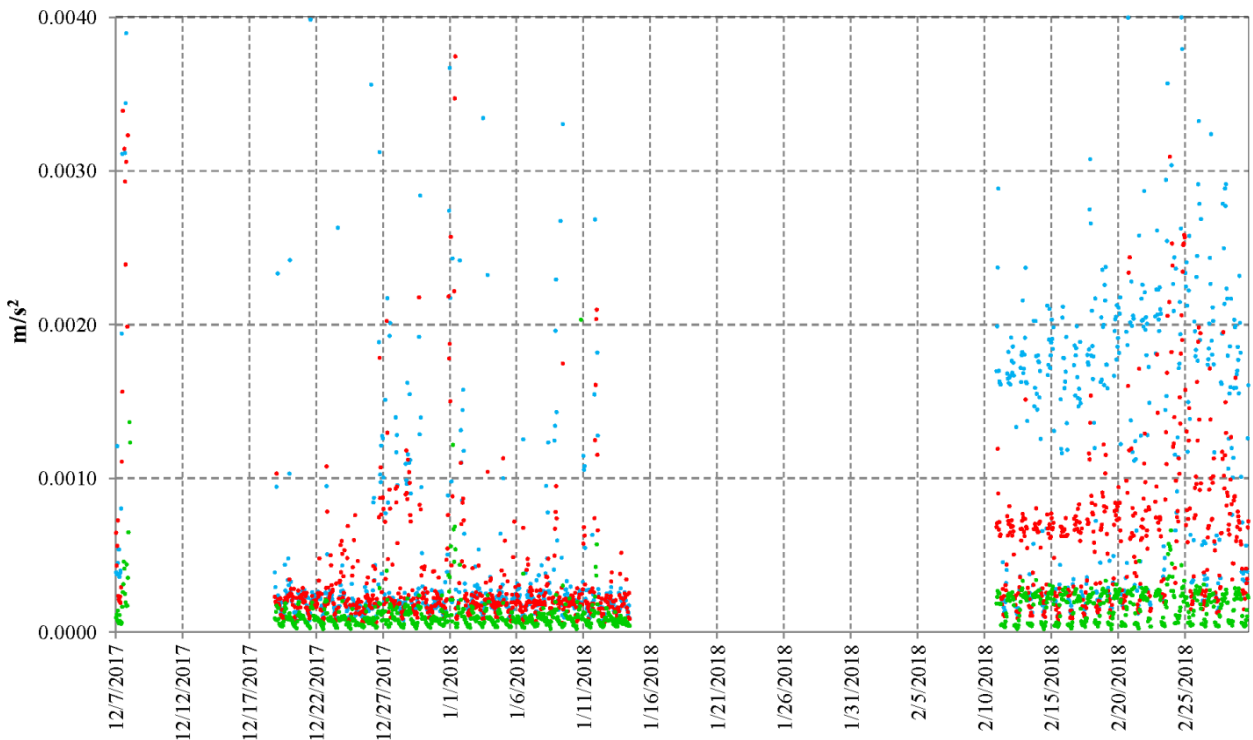

Figure 14. Maximum absolute values per hour of the accelerations recorded by AGI stations (S.2, +37 m) in the $x$ (blue dots), $y$ (red dots) and $z$ (green dots) directions.

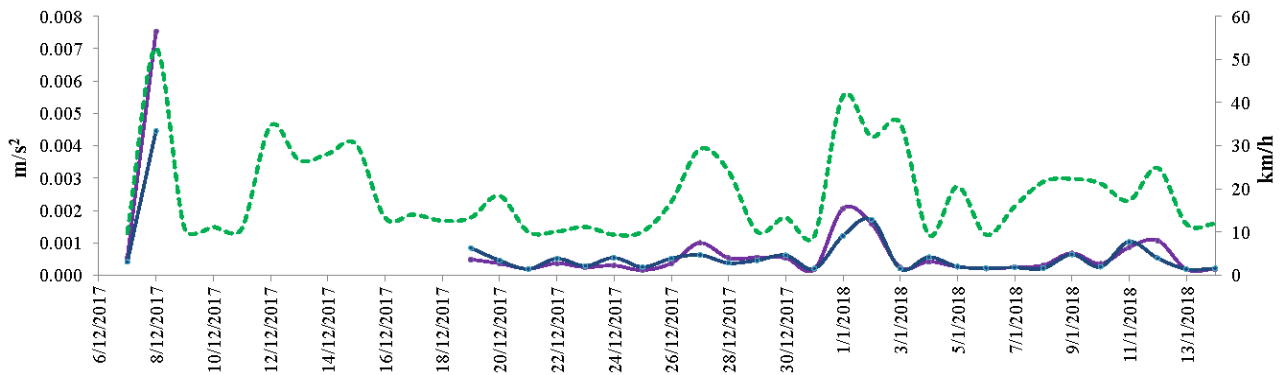

Figure 15. Daily average of the maximum hourly accelerations measured by S.2 vs. time along the *x* (blue) and *y* (red). directions. The dashed green line stands for the daily maximum wind speed.

Figures 16 and 17 provide information on the tower's vibrations over time in the band [0, 25] Hz. In particular, Figure 16 shows the spectrogram of the signal recorded by S. 942, at the base of the tower, in the period 18 November to 26 November 2017, for the three channels *x* (up), *y* and *z* (bottom). The tower's natural frequencies are clearly visible in the band [0, 10] Hz (horizontal lines), as well as the 3 Hz frequency of the building adjacent to the tower, along the *x* direction. As a result of anthropogenic activity, the power frequency increases in the entire band during the daytime, in particular during week-ends (18-19 November, 25-26 November). The three-dimensional spectrogram of channel *x* plotted in Figure 17 also highlights other anthropogenic activity between 20 and 25 Hz, which could be attributed to traffic in the historic centre. At about 12:30 (UTC) on 19 November, there is a clear frequency peak, which can be attributed to the M4.4 Parma earthquake reported in Table 3.

Many seismic events have been detected on the tower during the monitoring period. Table 3 shows the main earthquakes recorded, while Table 4 reports the teleseismic events identified in the signals. Figure 18 shows the teleseismic record of the Peru earthquake, which occurred at 09:18.44 (UTC) with a magnitude of 7.1, and was recorded at about 09:33 (UTC) by the instruments on the tower. In particular, the figure shows the signals recorded by the SS20 seismic stations at different levels along the tower's height in the *x* direction. The same event was also recorded by the S.2 accelerometer. At such a considerable distance from the epicentre, the seismic signal loses its high frequency content and becomes recognizable only at low frequencies. The records in the Figure were obtained by filtering the signals via a Butterworth band-pass filter with lower cut-off frequency of 0.04 Hz and a higher cut-off cut at 1 Hz. It is worth noting that after the first arrival of the waves the event remains clearly recognizable in the signals for long time.

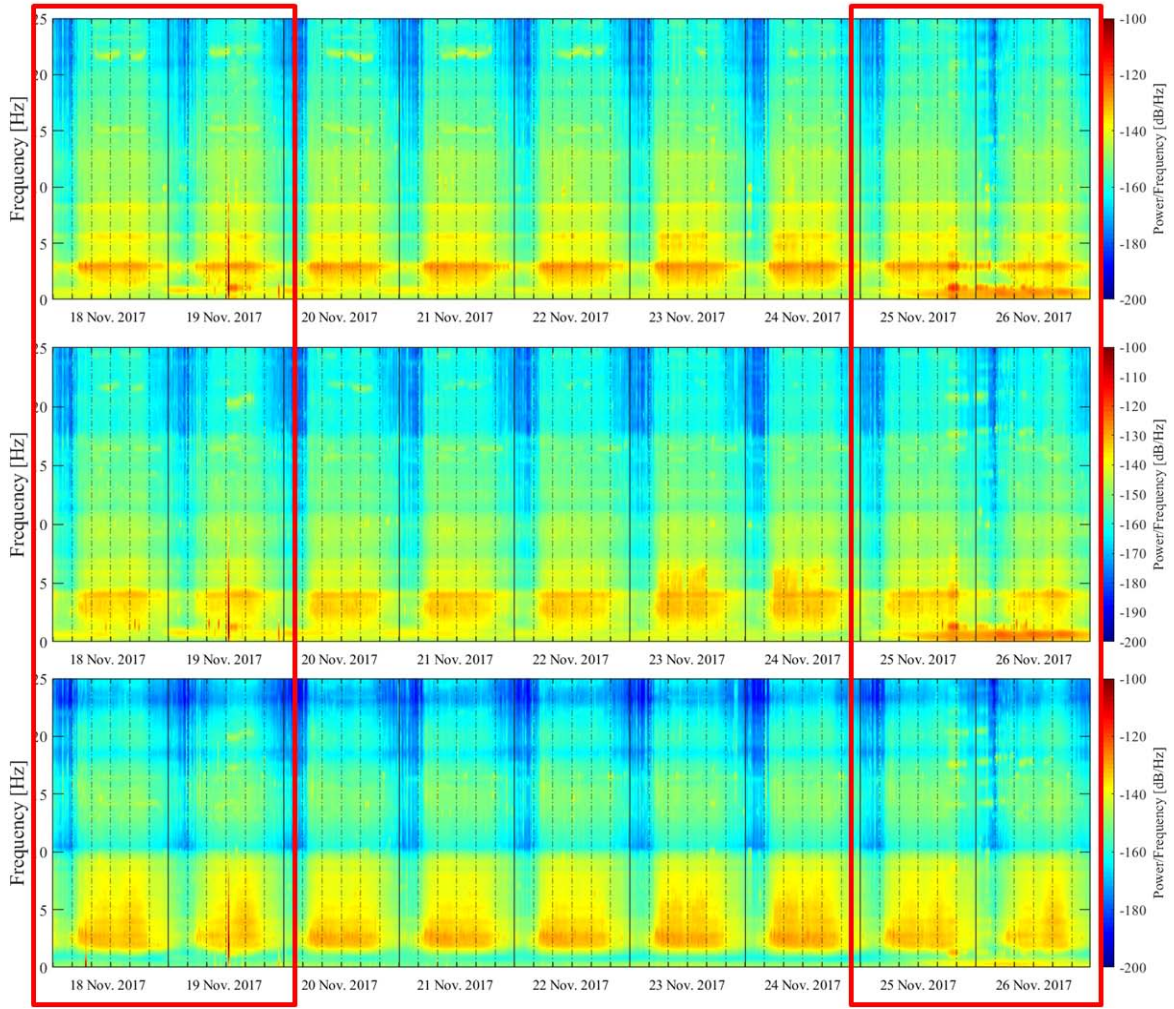

Figure 16. Spectrogram of the signal recorded by S.942 (+ 0,00 m) in the *x* (up), *y* and *z* (bottom) directions from 18 to 26 November 2017. Week-ends are highlighted by the red boxes.

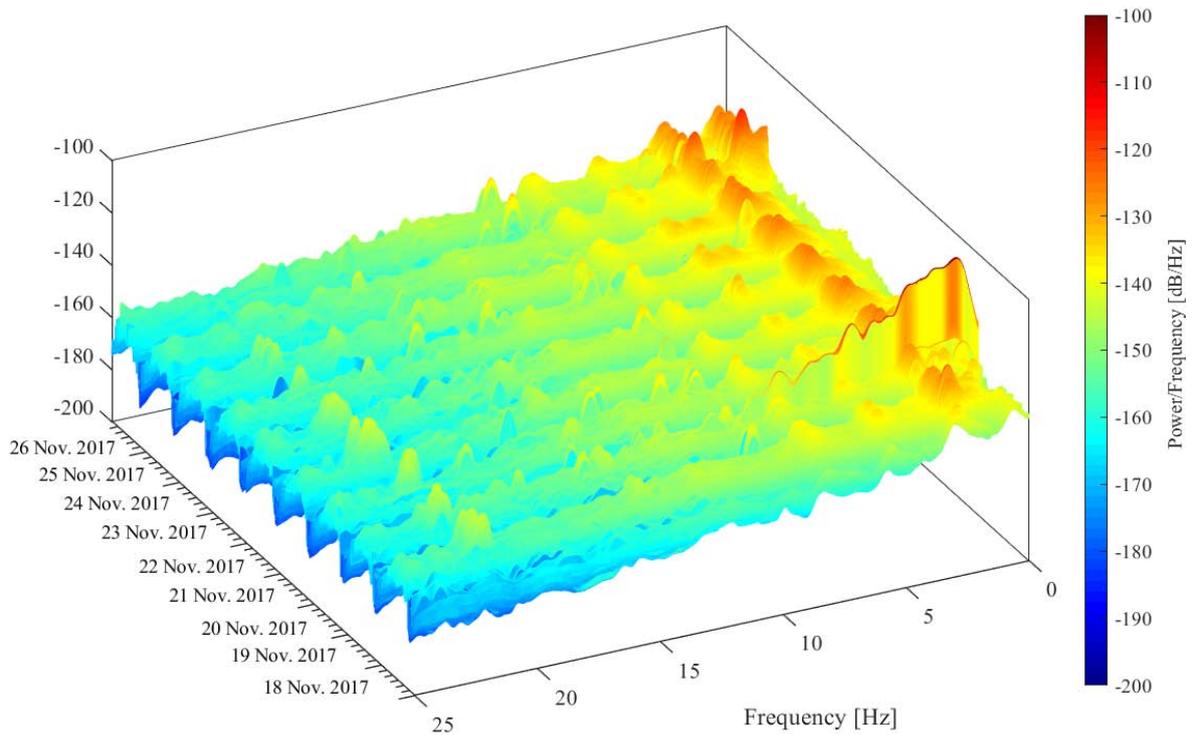

Figure 17. Three-dimensional spectrogram of the signal recorded by S.942 (+0,00 m) in the *x* direction from 18 to 26 November 2017.

| Date | Time (UTC) | Location | Magnitude |
|---|---|---|---|
| **11/19/17** | 12:10:12 | Parma | **3.3** |
| **11/19/17** | **12:37:44** | **Parma** | **4.4** |
| **12/03/17** | 23:34:11 | Amatrice | **4.0** |

Table 3. Earthquakes recorded on the Clock tower in the monitoring period.

| Date | Time (UTC) | Location | Magnitude |
|---|---|---|---|
| **11/17/17** | 22:34:21 | China | **6.4** |
| **11/19/17** | 09:25:50 | New Caledonia | **6.4** |
| **11/19/17** | 15:09:04 | New Caledonia | **6.6** |
| **11/19/17** | 22:43:31 | New Caledonia | **6.9** |
| **11/30/17** | 06:32:50 | Central Mid-Atlantic Ridge | **6.3** |
| **12/01/17** | 02:32:48 | Iran | **6.2** |
| **01/11/18** | 18.26.24 | Myanmar | **6.0** |
| **01/14/18** | **09:18.44** | **Peru** | **7.1** |
| **01/23/18** | 09:31:43 | United States (sea) | **7.6** |
| **01/24/18** | 10:51:19 | Japan | **6.4** |
| **01/25/18** | 01:15:59 | India | **6.0** |
| **01/25/18** | 02:10:37 | Russia | **6.3** |

Table 4. Teleseismic earthquakes recorded on the Clock tower in the monitoring period.

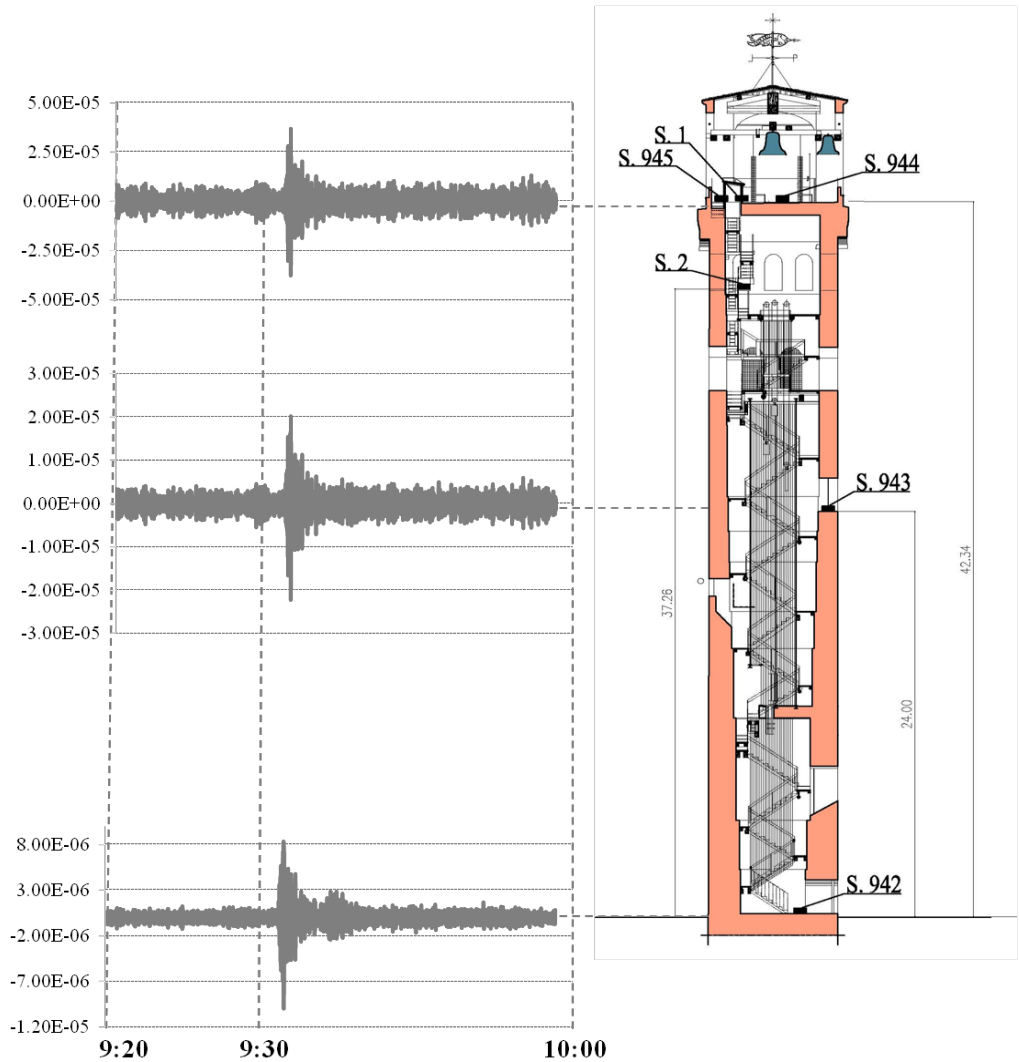

Figure 18. Signal of the Peru earthquake recorded on the tower in the *x* direction by S.942, S. 943, S.944 on 14 January 2018 from 9:33 (UTC) onward. Signal amplitude is in m/s$^2$.

Analyzing the low frequency content of the recorded signals can provide further information on movements of the tower. Figure 19 shows the signals recorded by the S.2 accelerometer from 11 to 14 January 2018 along the *x* (up) and *y* (bottom) directions. The signals in the figure were low-pass filtered with a cut-off frequency of 0.5 Hz, and clearly reveal a slow oscillation with period of about 24 hours. The same oscillations can be detected in the temperature signal (Figure 20) measured by the sensor inside the instrument. The temperature signal is however not in phase with the accelerations, thus demonstrating that the oscillations of the accelerometric signal are not a direct effect of temperature on the instrument. It is also worth noting that the signals along the *x* (North-South direction) and *y* (East-West direction) directions present a phase shift of 3 to 6 hours, and the *y* signal is delayed with respect to that along *x*. A possible interpretation of these oscillations lies in the daily temperature variations caused by exposure of the tower to the sun. As the exposure of the façades changes during the day, the tower appears to twist as shown in the figures. A measure of the displacements induced by this effect at the top of the tower can be easily obtained from the figures. In fact, for small oscillations, the modulus of the tangential acceleration expressed in *g* coincides with the inclination angle of the tower in radians. Thus, taking into account the actual position of the instrument (at +37 m), horizontal displacements turn out to be on the order of 3-4 mm for both directions. These daily movements appear to be the largest that the tower undergoes under normal conditions, as an effect of environmental factors.

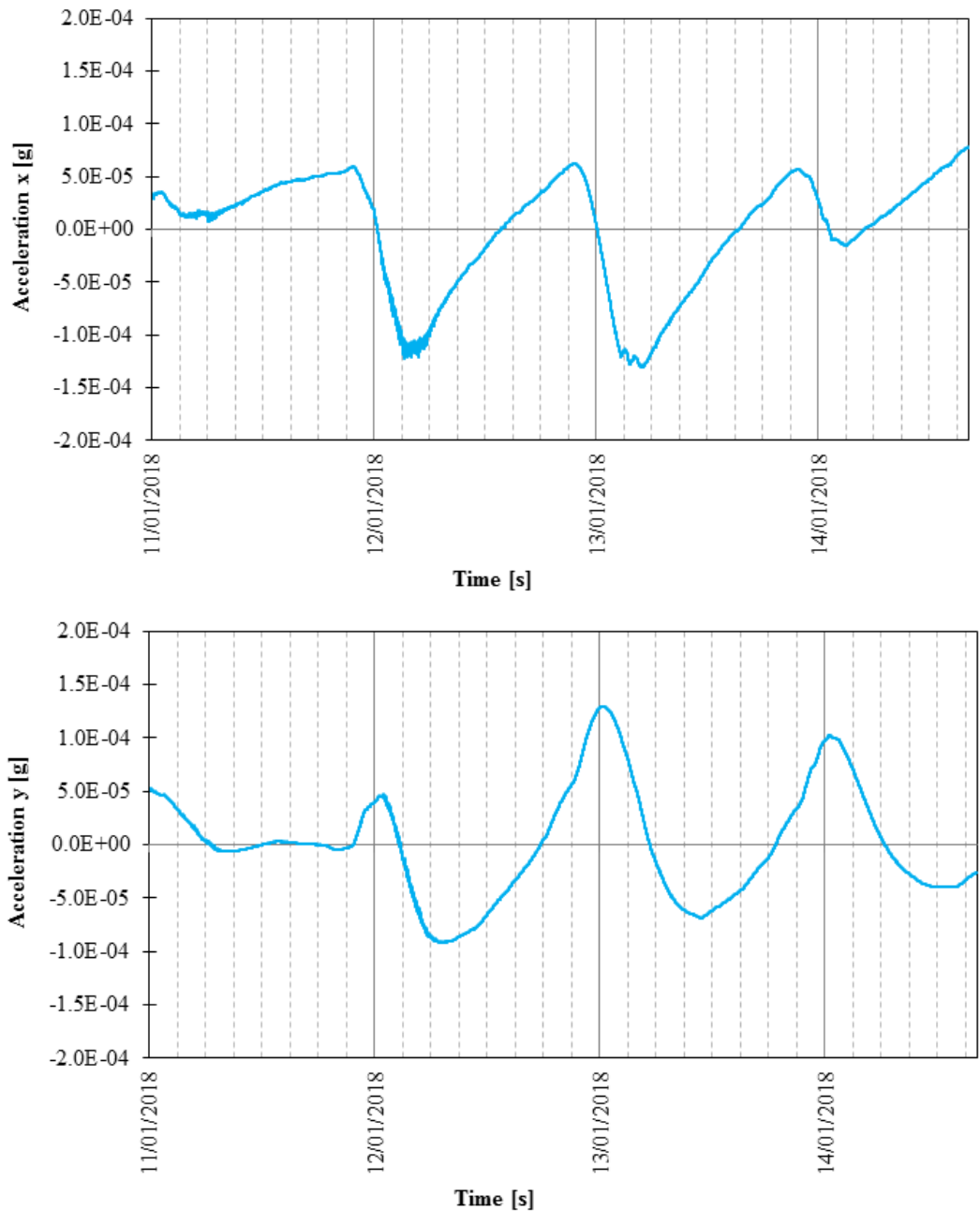

Figure 19. Low-pass filtered accelerations recorded by S.2 along the *x* and *y* (bottom) directions from 11 to 14 January 2018.

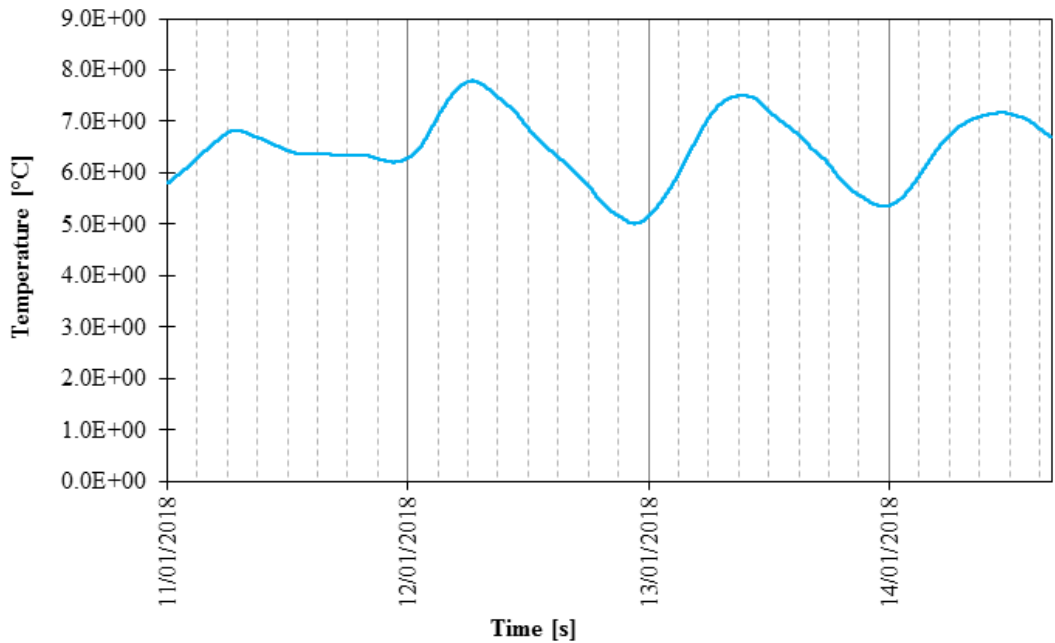

Figure 20. Temperatures recorded by the sensor installed inside S.2 from 11 to 14 January 2018.

## 4. Conclusions

The paper presents the results of an ambient vibration monitoring campaign conducted using high-sensitivity instrumentation on the Lucca Clock Tower from November 2017 to March 2018. In particular, four seismic stations provided by the INGV and two three-axial accelerometers developed by AGI S.r.l were installed on the tower. The combined use of two different kinds of measurement devices allowed for a comparison of the instruments' performance, as well as cross-validation of the results obtained for the Clock Tower. In addition, exploiting the different characteristics of seismometers and accelerometers made it possible both to explore the dynamic behaviour of the tower and highlight the effects of environmental vibrations over a wide range of frequency.

Analysis of the large dataset collected on the tower has yielded the following main results:
a) despite the low vibration levels, the main sources of vibration have been identified, as have the dynamic characteristics of the tower;
b) the chief factors influencing the tower's dynamic properties are: temperature, wind speed, ringing bells, crowd movements, traffic around the historic centre, and micro-tremors;
c) the instruments on the tower are able to clearly detect both earthquakes at epicentral distance of up to hundreds of kilometres and, in the low frequency range, teleseismic sequences;
d) day-to-day movements of the tower can be also deduced by analysing the low-frequency signals;
e) the simultaneous presence of the two instruments on the tower provides a detailed cross-validation of the results.

This study shows the potentials of continuous long-term monitoring in both investigating the influence of the surrounding environment on heritage structures and enhancing our knowledge of their structural health.